\documentclass{iaus}
\usepackage{graphicx}

\title[SiO maser intensity ratios] 
{SiO maser observations of a wide dust-temperature range sample}

\author[J. Nakashima \& S. Deguchi]   
{Jun-ichi Nakashima$^1$ \and Shuji Deguchi$^2$}

\affiliation{$^1$Academia Sinica Institute of Astronomy and Astrophysics, 
P.O. Box 23-141, Taipei 10617, Taiwan \break email: junichi@asiaa.sinica.edu.tw \\[\affilskip]
$^2$Nobeyama Radio Observatory, National Astronomical Observatory, 
Minamimaki, Minamisaku, Nagano 384-1305, Japan}

\pubyear{2007}
\volume{242}  
\pagerange{001--005}
\date{version 04/02/2007}
\jname{Astrophysical Masers and Their Environments}
\editors{J. Chapman \& W. Baan, eds.}
\begin{document}

\maketitle

\begin{abstract}
We present the results of SiO line observations of a sample of known SiO maser sources covering a wide dust-temperature range. The aim of the present research is to investigate the causes of the correlation between infrared colors and SiO maser intensity ratios among different transition lines. We observed in total 75 SiO maser sources with the Nobeyama 45m telescope quasi-simultaneously in the SiO~$J=1$$-$0~$v=0$,~1,~2,~3,~4 and $J=2$$-$1~$v=1$,~2 lines. We also observed the sample in the $^{29}$SiO~$J=1$$-$0~$v=0$ and $J=2$$-$1~$v=0$, and $^{30}$SiO~$J=1$$-$0~$v=0$ lines, and the H$_2$O~6$_{1,6}$$-$5$_{2,3}$ line. As reported in previous papers, we confirmed that the intensity ratios of the SiO~$J=1$$-$0~$v=2$ to $v=1$ lines clearly correlate with infrared colors. In addition, we found possible correlation between infrared colors and the intensity ratios of the SiO~$J=1$$-$0~$v=3$ to $v=1\&2$ lines. 
\end{abstract}

\firstsection 
\section{Introduction}

An important problem in the studies on the SiO maser was that SiO maser sources ever known were considerably biased. Specifically, the dust (effective) temperature of known SiO maser sources, which was calculated from mid-infrared flux densities (such as the IRAS and MSX flux densities), was limited roughly in a range of 250~K~$\lesssim~T_{\rm dust}~\lesssim~2000$~K. This is because the previous SiO maser surveys have been limited to relatively warm dust-temperature ranges. Consequently a non-negligible number of potential SiO maser sources (especially with a low dust-temperature) have been slipped from the previous SiO maser surveys. 

\cite{nym93} first realized the importance of SiO maser sources exhibiting a low dust-temperature. They investigated how SiO maser emission behaves in a low dust-temperature range by observing OH/IR stars in the SiO $J=1$$-$0~$v=1\&2$ and $J=$2$-$1~$v=1$ lines. The OH/IR stars often exhibit a low dust-temperature less than $T_{\rm dust}=250$~K. In their observation cold objects clearly show a larger intensity ratio of the SiO~$J=1$$-$0~$v=2$ to $v=1$ lines. Both collisional and radiative schemes cannot fully explain this observational properties of the SiO masers (\cite[Bujarrabal 1994]{buj94,doe95}; \cite[Deol et al. 1995]{doe95}). \cite{nym93} suggested that an infrared H$_2$O line ($11_{6,6}$~$\nu_{2}=1$$\rightarrow$$12_{7,5}$~$\nu_{2}=0$) overlapping with the SiO~$J=0$~$v=1$$\rightarrow$$J=1$ $v=2$ transition might play an important role. However, in early 1990s the number of cold SiO maser sources (like OH/IR stars) was quite limited, and it was difficult to statistically investigate the relation between infrared colors and intensity ratios of SiO maser lines.

\cite{nak03b} recently extended the Nyman's study by surveying the SiO maser emission in cold, dusty IRAS sources exhibiting low dust temperature less than 250~K. They found roughly 40 new SiO maser sources in the cold dusty objects, and in conjunction with the results of another SiO maser survey of relatively warm IRAS objects (\cite[Nakashima \& Deguchi 2003a]{nak03a}) they clearly demonstrated that the intensity ratio of the SiO~$J=1$$-$0~$v=2$ to $v=1$ lines increases in inversely proportional to the dust temperature. \cite{nak03b} again suggested that the overlap line of H$_2$O might explain this correlation if the overlap line becomes stronger with decrease of the dust temperature. To consider further this problems, we need to confirm whether properties of the SiO lines other than $J=1$$-$0~$v=1$ and 2 lines are consistent with the existence of the H$_2$O overlap line. In this contributed paper we present the result of quasi-simultaneous observations in the multiple different SiO rotational lines with the Nobeyama 45m telescope. The main aim of the observation is to check the behavior of SiO maser intensity ratios including lines other than the $J=1$$-$0~$v=1$ and 2 lines.

\section{Observations and Results}

The observing targets were selected from Nakashima \& Deguchi(2003a, b) and the Nobeyama SiO maser source catalog (Gorny et al. in preparation) in terms of the IRAS colors and flux densities. The targets are distributed roughly in the right ascension range between $18^{\rm h}$ and $22^{\rm h}$, because the cold SiO maser sources found by \cite{nak03b} are distributed roughly in this range. We selected the observing targets basically in order of the brightness at $\lambda =12$ $\mu$m, but we also paid attention to the source distribution in the IRAS two-color diagram so that the observing targets continuously cover the entire color range. 

SiO line observations with the Nobeyama 45m telescope were made in two separated periods: May 11--19, 2004 and February 15--19, 2006. In the first period we observed, in total, 38 objects. The observed SiO transitions in the first period were $J=1$$-$0~$v=$1,~2,~3 and $J=2$$-$1~$v=$1,~2. We also observed in the $^{29}$SiO~$J=1$$-$0~$v=$0 and $J=2$$-$1~$v=$0 lines. In addition, we observed 27 objects in the H$_2$O maser line at 22 GHz ($6_{1,6}$$-$5$_{2,3}$) as a backup observation under rainy/heavy cloudy condition. In the second period we observed, in total, 53 objects. The observed transitions in the second period were SiO $J=1$$-$0~$v=$0,~1,~2,~3,~4, $^{29}$SiO $J=1$$-$0~$v=$0 and $^{30}$SiO $J=1$$-$0~$v=$0. The technical details of the observations will be presented in our future paper (Nakashima \& Deguchi, in preparation).

\begin{figure}
\centerline{
\scalebox{0.8}{%
\includegraphics{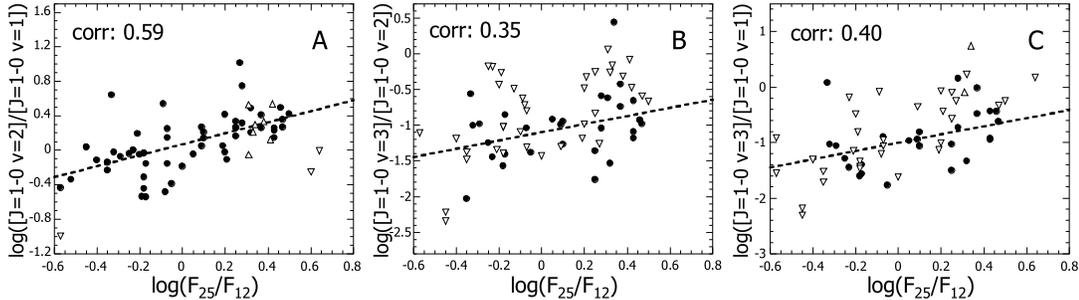}%
}
}
  \caption{Infrared colors versus intensity ratio of SiO maser lines. The horizontal axes represent infrared colors. $F_{25}$ and $F_{12}$ denote the IRAS flux densities at $\lambda=25$ and 12 $\mu$m, respectively. The filled dots ($\bullet$), upward triangles ($\bigtriangleup$) and downward triangles ($\bigtriangledown$) respectively represent the intensity ratios of the SiO maser lines, lower limits of the ratio and upper limits of the ratio. Correlation coefficients are given in the upper-left corners of each panel. The dashed lines are the results of least-square fitting of a first order polynomial.}\label{fig:ratio-color}
\end{figure}

In this paper, we focus on the properties of the SiO~$J=1$$-$0~$v=1$,~2~and~3 lines, in which we detected an enough number of objects for statistical analysis. Figure 1 shows the relations between infrared colors and intensity ratios among the SiO maser lines. The line intensities used to calculate the intensity ratios are velocity-integrated intensities. In the panel A of Figure 1 we can clearly confirm the positive correlation between the $\log(F_{25}/F_{12})$ color and the intensity ratio of the SiO~$J=1$$-$0~$v=2$ to $v=1$ lines as reported by \cite{nak03b}. Interestingly, the intensity ratios of the $J=1$$-$0 $v=3$ to $v=2$ lines and of the $J=1$$-$0~$v=3$ to $v=1$ lines seem to also correlate with the $\log(F_{25}/F_{12})$ color (see, panels B and C) even though the correlation coefficients are slightly smaller than that of panel A.

\begin{figure}
\centerline{
\scalebox{0.75}{%
\includegraphics{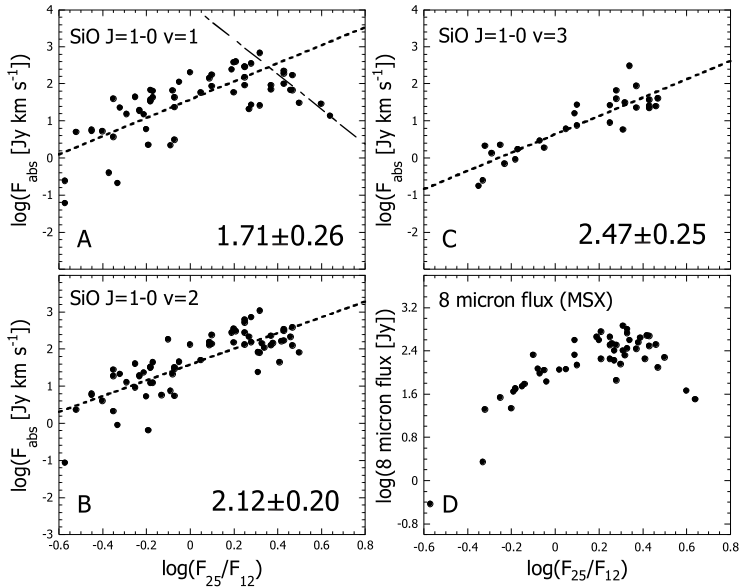}%
}
}
  \caption{[{\it Panels A, B and C}]: Relation between infrared colors and absolute intensity of SiO maser lines. The notation of the infrared colors is same with that used in Figure 1. The intensity of SiO maser lines is standardized at the distance of 1 kpc. The thick dashed lines represent the results of least-square-fitting of a first order polynomial. The inclinations of the fitted lines (thick dashed lines) are given at the lower-right corners of each panel with statistical uncertainty. In the panel A, only the data points below $\log(F_{25}/F_{12})=0.5$ were fitted by the polynomial. The data points above $\log(F_{25}/F_{12})=0.5$ are independently fitted by a first order polynomial, and the results of the fitting is given as the chain line. [{\it Panel D}]: Relation between 8$\mu$m absolute flux density and infrared color. The 8$\mu$m flux is standardized at the distance of 1 kpc.}\label{fig:int-color}
\end{figure}

Figure 2 shows the relations between infrared colors and absolute intensities of the SiO maser lines. The intensity of the SiO maser lines is standardized at the distance of 1 kpc using the luminosity distances. The panels A, B and C of Figure 2 show the relations between the $\log(F_{25}/ F_{12})$ color and the absolute intensity of the SiO $J=1$$-$0~$v=1$, 2 and 3 lines. A notable feature seen in these panels is that the SiO maser absolute intensities undoubtedly correlate with the $\log(F_{25}/ F_{12})$ color. Another clear feature is that the higher the vibrational transitions, the steeper the inclination of the dashed lines representing the results of least-square-fitting of a first order polynomial. This tendency is consistent with the correlation seen in Figure 1. In the panel A in Figure 2, the values of the absolute intensity of SiO maser emission seem to maximize at $\log(F_{25}/ F_{12}) \sim 0.5$, and the values tend to decrease with increase of the color in the red region above $\log(F_{25}/ F_{12})=0.5$. The $\log(F_{25}/ F_{12})$ color of 0.5 corresponds the boundary between distributions of AGB and post-AGB stars in the $\log(F_{25}/ F_{12})$ color. In fact, the panel A in Figure 1 (and Figure 8 in \cite[Nakashima \& Deguchi 2003b]{nak03b}) shows a sudden change of the feature at $\log(F_{25}/ F_{12}) \sim 0.5$. No such change is seen in the panels B and C in Figure 2, simply because the SiO $J=1$$-$0~$v=2$ and 3 lines have not been detected above $\log(F_{25}/ F_{12})=0.5$ in the present observations. 

A possible reason for the correlation seen in panels A, B and C in Figure 2 is that the energy input to the SiO maser region increases with the infrared colors. To confirm this possibility, in panel D in Figure 2 we plotted the 8$\mu$m flux densities as a function of the infrared colors. The values of the 8$\mu$m flux densities were taken from the MSX point source catalog. If we rely on the radiative scheme the 8$\mu$m flux should well represent the energy input to the SiO maser region, because the $\lambda=8\mu$m corresponds to the $\Delta v=1$ SiO transition (\cite[e.g., Deguchi \& Iguchi 1976]{deg76}). In panel D in Figure 2 the 8$\mu$m flux densities are standardized at the distance of 1kpc using the luminosity distances. The distribution of the data points seen in panel D is, in fact, strikingly similar with those seen in panels A, B and C, supporting that the 8$\mu$m flux tightly correlates with the SiO maser intensity as suggested by \cite{buj87}.

\section{Discussion}
In this section we discuss the possible explanation for the correlation between infrared colors and SiO maser intensity ratios among the $v=1$, 2 and 3 lines at 43 GHz. One possible explanation is to introduce the overlap line of H$_2$O ($11_{6,6}$ $\nu_{2}=1$~$\rightarrow$~$12_{7,5}$ $\nu_{2}=0$), which has been first suggested by \cite{olo81} to explain the anomalous, weak intensity of the SiO $J=2$$-$1~$v=2$ line in oxygen-rich (O-rich) stars. This H$_2$O line overlaps with the SiO $J=0$~$v=1$$\rightarrow$$J=1$~$v=2$ transition with a velocity difference of 1~km s$^{-1}$. With this line overlap, the $J=1$~$v=2$ level is overpopulated, and the weakness of the SiO $J=2$$-$1~$v=2$ line is explained by this overpopulation. The overpopulation at the $J=1$~$v=2$ level is also consistent with the strong intensity of the $J=1$$-$0~$v=2$ line. Thus, the correlation between the infrared colors and the intensity ratio of the SiO $J=1$$-$0~$v=2$ to $v=1$ lines may be explained if this overlap line of H$_2$O becomes stronger with increase of the infrared colors. One problem in this interpretation is that the intensity ratios of the SiO $J=1$$-$0 $v=3$ to $v=1\&2$ lines cannot be explained only by the H$_2$O $11_{6,6}$~$\nu_{2}=1$~$\rightarrow$~$12_{7,5}$ $\nu_{2}=0$ line. However \cite{cho07} recently reported an interesting detection of the SiO $J=2$$-$1~$v=3$ line toward an S-type star, $\chi$ Cyg. They also confirmed that the SiO $J=2$$-$1 $v=3$ line is weak in O-rich stars. The S-type stars have almost same amount of oxygen and carbon atoms in their envelopes, and consequently they have few H$_2$O molecules in the envelopes. These results potentially suggest that another overlap line of H$_2$O affects on the population distribution of SiO in O-rich stars, and \cite{cho07} have suggested that the H$_2$O $5_{0,5}$~$\nu_{2}=2$$\rightarrow$$6_{3,4}$~$\nu_{2}=1$ line overlapping with the SiO $J=0$ $v=2$$\rightarrow$$J=1$ $v=3$ line (with a velocity difference of about 1.5 km s$^{-1}$) acts on the population distribution of SiO. Thus, if both H$_2$O $11_{6,6}$~$\nu_{2}=1$$\rightarrow$$12_{7,5}$~$\nu_{2}=0$ and $5_{0,5}$~$\nu_{2}=2$$\rightarrow$$6_{3,4}$~$\nu_{2}=1$ lines becomes stronger with increase of infrared colors, all correlations between infrared colors and the SiO maser intensity ratios among the $J=1$$-$0 $v=1$, 2 and 3 lines might be explained. The line intensity of the H$_2$O $5_{0,5}$~$\nu_{2}=2$$\rightarrow$$6_{3,4}$~$\nu_{2}=1$ line is usually weaker than that of the $11_{6,6}$~$\nu_{2}=1$$\rightarrow$$12_{7,5}$~$\nu_{2}=0$line. This fact also seems to be consistent with the relatively weak intensity of the SiO $J=1$$-$0 $v=3$ line.

However, there are some other problems on the explanation with the overlap line of H$_2$O. First, we have to explain how the H$_2$O infrared lines overlapping with the SiO lines become stronger with increase of infrared colors. The relative abundance of H$_2$O molecules possibly increases with infrared colors, but this is not conclusive. Second, the correlation between infrared colors and the intensity ratios of the SiO $J=1$$-$0 $v=2$ to $v=1$ lines might be explained without the overlap line of H$_2$O. In the envelopes of very cold objects, strong 8$\mu$m emission comes from every direction to the SiO masing region, causing ineffective pumping through the SiO $\Delta v=1$ transition. On the other hand, 4$\mu$m emission corresponding to the SiO $\Delta v=2$ transition is more effectively pump the SiO population instead of the 8$\mu$m. These processes might explain the correlation seen in Figure 3 (\cite[e.g., Doel et al. 1995]{doe95}). Third, a recent theoretical calculation predicted that if we introduce the overlap line of H$_2$O the spatial distribution of the maser spots cannot be theoretically reproduced (\cite[Soria-Ruiz et al. 2004]{sor04}). Thus, this problem will be remain controversial for some more time.

\section{Summary}
In this research we observed 75 known SiO maser sources quasi-simultaneously in the SiO $J=1$$-$0, $v=0$, 1, 2, 3 and 4 lines, SiO $J=2$$-$1~$v=1$ and 2, $^{29}$SiO~$J=1$$-$0~$v=0$ and $J=2$$-$1~$v=0$, and $^{30}$SiO~$J=1$$-$0~$v=0$ lines. We also observed the targets in the H$_2$O~6$_{1,6}$$-$5$_{2,3}$ line under rainy/heavy cloudy condition. The sample continuously covers a very wide dust-temperature range from 150~K to 2000~K. The correlation between infrared colors and the intensity ratio of the SiO $J=1$$-$0 $v=2$ to $v=1$ lines is confirmed as reported by \cite{nak03b}. The intensity rations of SiO $J=1$$-$0 $v=3$ to $v=1$\&2 lines possibly correlate with infrared colors. The overlap lines of H$_2$O might explain the correlations between the infrared colors and the SiO maser intensity ratios among the $J=1$$-$0~$v=1$, 2 and 3 lines, although there are alternative ways to interpret the phenomena.

\begin{acknowledgments}
The present research has been supported by the Academia Sinica Institute of Astronomy \& Astrophysics in Taiwan. \end{acknowledgments}


\begin{discussion}

\discuss{Elitzur}{In our recent work, infrared intensity and colors are uniquely determined by the optical depth of the dust shell. SiO maser intensity is also correlated to the optical depth. Therefore, what you are finding is somehow related to the effect of the activity of the photosphere such as the variation of mass loss rates.}

\discuss{Nakashima}{Thanks for useful comments. (We took account of Elitzur's comments in the text.)}

\end{discussion}

\end{document}